\documentclass[prd,twocolumn,preprintnumbers,amsmath,amssymb,showpacs]{revtex4}

\usepackage{graphicx,bm}

\makeatletter
\def\graphicscale{\twocolumn@sw{0.3}{0.4}}
\def\graphicthreescale{\twocolumn@sw{0.3}{0.4}}

\begin{document}

\title{

Criticality of O($N$) symmetric models in the presence of discrete gauge symmetries

}

\author{Andrea Pelissetto}
\affiliation{Dipartimento di Fisica dell'Universit\`a di Roma ``La Sapienza"
        and INFN, Sezione di Roma I, I-00185 Roma, Italy}

\author{Antonio Tripodo, Ettore Vicari}
\affiliation{Dipartimento di Fisica dell'Universit\`a di Pisa
        and INFN, Sezione di Pisa, I-56127 Pisa, Italy}

\date{\today}

\begin{abstract}

We investigate the critical properties of the three-dimensional (3D)
antiferromagnetic RP$^{N-1}$ model, which is characterized by a global
O($N$) symmetry and a discrete ${\mathbb Z}_2$ gauge symmetry.  We
perform a field-theoretical analysis using the Landau-Ginzburg-Wilson
(LGW) approach and a numerical Monte Carlo study.  The LGW
field-theoretical results are obtained by high-order perturbative
analyses of the renormalization-group (RG) flow of the most general
$\Phi^4$ theory with the same global symmetry as the model, assuming a
gauge-invariant order-parameter field.  For $N=4$ no stable fixed
point is found, implying that any transition must necessarily be of
first order. This is contradicted by the numerical results that
provide strong evidence for a continuous transition.  This suggests
that gauge modes are not always irrelevant, as assumed by the LGW
approach, but they may play an important role to determine the actual
critical dynamics at the phase transition of O($N$) symmetric models
with a discrete ${\mathbb Z}_2$ gauge symmetry.
\end{abstract}



\maketitle


\section{Introduction}
\label{intro}

In the framework of the renormalization-group (RG) theory of critical
phenomena, the Landau-Ginzburg-Wilson (LGW) field-theoretical
approach~\cite{Landau-book,WK-74,Fisher-75,Ma-book,ZJ-book,PV-02}
provides accurate descriptions of continuous phase transitions in many
physical systems. The starting point is the identification of the
order parameter associated with the critical modes and of the
symmetry-breaking pattern characterizing the transition.  Then, one
considers the corresponding LGW $\Phi^4$ field theory, which is the
most general fourth-order polynomial theory of the order-parameter
field with the same symmetries as the original model.  The analysis of
the corresponding RG flow provides the universal features of the
critical behavior.

When the statistical system under investigation presents also a gauge
symmetry, the traditional LGW approach generally assumes a
gauge-invariant order parameter.  Then the nature of the critical
behavior is inferred from the RG flow of the $\Phi^4$ theory that is
invariant under the global symmetries of the original model. In this
approach the gauge degrees of freedom are effectively integrated out,
assuming that they do not play a significant role at the phase
transition.  However, as pointed out in Ref.~\cite{PTV-17}, this
approach fails for some phase transitions.  In the case of the
three-dimensional (3D) CP$^{N-1}$ models, characterized by a global
U($N$) symmetry and a U(1) gauge symmetry, the predictions of the
corresponding LGW theories are not consistent with the critical
behavior observed in a variety of models with the same gauge and
global symmetries \cite{PTV-17,DPV-15,NCSOS-11}, with only a few
exceptions.

In this paper we again discuss this issue, checking whether the
above-mentioned LGW approach also fails in the presence of discrete
gauge symmetries.  For this purpose, we consider 3D RP$^{N-1}$ models
that are characterized by a global O($N$) symmetry and a discrete
${\mathbb Z}_2$ gauge symmetry.  In particular, we consider the
antiferromagnetic RP$^{N-1}$ (ARP$^{N-1}$) model, which undergoes a
continuous transition for both $N=2$ and $N=3$ \cite{FMSTV-05}.  To
apply the standard LGW approach, we identify a local gauge-invariant
order-parameter field, that belongs to the spin-2 representation of
the O($N$) symmetry, and construct the corresponding O($N$)-symmetric
LGW $\Phi^4$ theory.  For $N=2,3$ this theory gives results that are
in full agreement with numerical investigations \cite{FMSTV-05}. We
extend here the analysis to the case $N=4$. We analyze the RG flow in
the LGW theory, finding no evidence of fixed points. Thus, the LGW
approach predicts the absence of continuous transitions for such value
of $N$. This prediction is, however, contradicted by numerical Monte
Carlo (MC) results.  A finite-size scaling (FSS) of data on lattices
of size up to $L=100$ gives a compelling evidence for a second-order
transition. Therefore, also in the case of a discrete gauge symmetry,
the LGW approach with a gauge-invariant order parameter may fail.
This provides a further evidence that LGW $\Phi^4$ theories
constructed using a gauge-invariant order-parameter field, thus
integrating out the gauge modes, do not generally capture the relevant
features of the critical dynamics.

The paper is organized as follows. In Sec.~\ref{arpcp} we construct
the LGW theory which is expected to describe the critical modes at the
continuous transitions of ARP$^{N-1}$ models, assuming a staggered
gauge-invariant order parameter.  In Sec.~\ref{hopa} we determine the
RG flow for $N= 4$, using high-order field-theoretical perturbative
series.  In Sec.~\ref{arp3res} we study numerically the nature of the
critical behavior of the ARP$^3$ model.  Finally, in Sec.~\ref{conclu}
we draw our conclusions.  The perturbative series and a discussion of
their large-order behavior are reported in the appendices.

\section{LGW theories for the ARP$^{N-1}$ models}
\label{arpcp}

In this section we derive the LGW theories associated with the
ARP$^{N-1}$ models, emphasizing the main assumptions and/or
hypotheses.  The effective LGW theory is generally constructed using
global properties such as the nature of the order parameter, the
symmetry of its critical modes, and the symmetry-breaking pattern.

RP$^{N-1}$ models are defined by the Hamiltonian
\begin{equation}
H_{\rm RP} = J \sum_{\langle {\bm x}{\bm y} \rangle} 
| \bm{s}_{\bm x} \cdot {\bm s}_{\bm y} |^2,
\label{hrpn}
\end{equation}
where the sum is over the nearest-neighbor sites ${\langle {\bm x}{\bm
    y} \rangle}$ of a cubic lattice, ${\bm s}_{\bm x}$ are
$N$-component real vectors satisfying $\bm{s}_{\bm x} \cdot {\bm
  s}_{\bm x}=1$. The model is ferromagnetic for $J<0$,
antiferromagnetic for $J > 0$.  RP$^{N-1}$ models present a global
O($N$) symmetry and a local ${\mathbb Z}_2$ gauge symmetry
(independent changes of the sign for each site variable).

Let us assume that the critical modes are effectively represented by
local gauge-invariant variables, which may be identified as the
gauge-invariant site variable
\begin{equation}
P_{\bm x}^{ab} = s_{\bm x}^a s_{\bm x}^b - {1\over N} \delta^{ab},
\label{pdef}
\end{equation}
which is a symmetric real and traceless $N\times N$ matrix.  It
transforms as
\begin{equation}
P_{{\bm x}} \to O^\dagger P_{{\bm x}} O,
\label{symmetry-O(N)}
\end{equation}
under global O($N$) transformations.  The next step to construct the
LGW Hamiltonian requires the identification of the order parameter of
the transition.

In the case of ferromagnetic models, i.e. when $J<0$, the
order-parameter field $\Phi^{ab}({\bm x})$ can be formally related to
a spatial average of the site variable (\ref{pdef}) over a large but
finite lattice domain. Then, the corresponding LGW field theory is
obtained by considering the most general fourth-order polynomial in
$\Phi$ consistent with the O($N$) symmetry (\ref{symmetry-O(N)}):
\begin{eqnarray}
{\cal H} &=& {\rm Tr} (\partial_\mu \Phi)^2 
+ r \,{\rm Tr} \,\Phi^2 +  w_0 \,{\rm tr} \,\Phi^3 \label{hlg}\\
&& +  \,{u_0\over 4}\, ({\rm Tr} \,\Phi^2)^2  + {v_0\over 4}\, {\rm Tr}\, \Phi^4 .
\nonumber
\end{eqnarray}
For $N=2$, the cubic term vanishes and the two quartic terms are
equivalent.  Therefore, one recovers the O(2)-symmetric LGW theory,
consistently with the equivalence between the RP$^1$ and the XY
model. For $N\ge 3$, the cubic term is generally expected to be
present. This is usually considered as the indication that phase
transitions of systems sharing the same global properties are of first
order, as one can easily infer using mean-field arguments.

In the case of antiferromagnetic interactions ($J>0$), the minimum of
the Hamiltonian (\ref{hrpn}) is locally realized by taking ${\bm
  s}_{\bm x}\cdot {\bm s}_{\bm y} = 0$ for any pair of
nearest-neighbor sites ${\langle {\bm x}{\bm y} \rangle}$.  Thus, at
variance with the ferromagnetic case, the antiferromagnetic
interactions give rise to a breaking of translational invariance in
the low-temperature phase.  Hence, we may assume that the critical
modes are related with the staggered site variable
\begin{equation}
A_{\bm x}^{ab}\equiv p_{\bm x} P_{\bm x}^{ab},
\label{aiab}
\end{equation}
where $P_{\bm x}^{ab}$ is defined in Eq.~(\ref{pdef}), and $p_{\bm x}$
is the parity of the site ${\bm x} \equiv (x_1,x_2,x_3)$ defined by
$p_{\bm x} = (-1)^{\sum_k x_k}$.  The corresponding order parameter
should be its spatial average
\begin{equation}
  M^{ab} = \sum_{\bm x} A_{\bm x}^{ab},
  \label{mdef}
  \end{equation}
which is a symmetric and traceless matrix. Moreover it changes sign
under translations of one site which exchange the two sublattices.
Then, as usual, in order to construct the LGW model, we replace $A$
with a local variable $\Phi$ as fundamental variable (essentially, one
may imagine that $\Phi$ is defined as $M$, but now the summation
extends only over a large, but finite, cubic sublattice).  Then, the
corresponding LGW theory is obtained by writing down the most general
fourth-order polynomial that is invariant under O($N$) transformations
and under the global ${\mathbb Z}_2$ transformation $\Phi \to -\Phi$,
i.e.~\cite{ACFJMRT-05}
\begin{eqnarray}
{\cal H}_a = {\rm Tr} (\partial_\mu \Phi)^2 + r \,{\rm Tr} \,\Phi^2 +
{u_0\over 4} \, ({\rm Tr} \,\Phi^2)^2 + {v_0 \over 4} \, {\rm Tr}\,
\Phi^4 .
\label{ahlg} 
\end{eqnarray}

Since any $2\times 2$ and $3\times3$ traceless 
symmetric matrix $\Phi$ satisfies
\begin{equation}
{\rm Tr}\, \Phi^4 = {1\over 2} ({\rm Tr} \,\Phi^2)^2,
\label{eq2e3}
\end{equation}
the two quartic terms of the Hamiltonian (\ref{ahlg}) are equivalent
for $N=2$ and $N=3$. Therefore the $N=2$ and $N=3$ $\Phi^4$ theories
(\ref{ahlg}) can be exactly mapped onto the O(2) and O(5) symmetric
$\Phi^4$ vector theories, respectively.  This implies that the
continuous transition of the ARP$^1$ and ARP$^2$ models belong to the
O(2) and O(5) vector universality classes, respectively.

Note that, in the case of the ARP$^2$ model, this scenario entails an
enlargement of the global O(3) symmetry at the critical point, because
the O(5) symmetry is a feature of its LGW theory only, i.e., of the
expansion up to fourth powers of $\Phi$. Indeed, one can easily check
that the sixth-order terms, such as ${\rm Tr}\,\Phi^6$, allowed by the
global symmetries of the ARP$^2$ model do not share the O(5) symmetry.
Since these terms are RG irrelevant at the fixed point,
the contribution of the O(5)-breaking terms is suppressed at the
critical point.  Therefore, the critical point (more precisely, its
asymptotic critical behavior) shows a dynamic enlargement of the
symmetry. Thus, the critical modes of the ARP$^2$ model are associated
with the effective symmetry breaking O(5)$\to$O(4) at the transition
point, although the microscopic global symmetry is O(3).  This
prediction has been accurately verified by the numerical analyses
reported in Refs.~\cite{FMSTV-05,ACFJMRT-05,Carmona-etal-03}.

When $N\ge 4$ the LGW theory (\ref{ahlg}) cannot be simplified,
therefore one must keep both quartic terms.  The stability domain of
${\cal H}_a$ can be determined by studying the asymptotic large-field
behavior of the potential
\begin{equation}  
V(\Phi) = r \,{\rm Tr} \,\Phi^2 + {u_0 \over 4} \, ({\rm Tr}
\,\Phi^2)^2 + {v_0\over 4} \, {\rm Tr}\, \Phi^4 .
\label{vpsi}
\end{equation}
This analysis can be easily performed by noting that $V(\Phi)$ only
depends on the $N$ eigenvalues $\lambda_a$ of the symmetric matrix
$\Phi$, which satisfy the condition $\sum_a \lambda_a = 0$.  The
theory is stable if
\begin{eqnarray}
u_0 + b_N v_0 > 0, \qquad
b_N = {N^2 - 3 N  + 3\over N(N-1) } ,
\label{stabcond1}
\end{eqnarray}
and  if
\begin{eqnarray}
&u_0 + {1\over N} v_0 > 0  \quad & {\rm for}\; {\rm even }\;N,
\label{stabcond2}\\
&u_0 + c_N v_0 > 0  \quad & {\rm for}\; {\rm odd }\;N,\nonumber 
\end{eqnarray}
where 
\begin{equation}
c_N = {N^2 + 3\over N(N^2-1)}.
\label{stabcond3}
\end{equation}
Physical systems corresponding to the effective theory (\ref{ahlg})
with $u_0,v_0$ that do not satisfy these constraints are expected to
undergo a first-order phase transition.

The analysis of the minima of the potential $V(\Phi)$ for $r<0$ gives
us information on the symmetry-breaking patterns. For $v_0<0$, the
absolute minimum of $V(\Phi)$ is realized by configurations with $\Phi
= O\Phi_{\rm min} O^\dagger$ and
\begin{equation}
\Phi_{\rm min} \sim
\left( 
\begin{array}{l@{\ \ }l@{\ \ }}
I_{N-1}& \quad 0 \\
\;\;0 & -(N-1) \\
\end{array}
\right), 
\label{psivp}
\end{equation}
where $I_{n}$ indicates the $n\times n$ identity matrix and 
$O$ is an orthogonal matrix.  This gives
rise to the symmetry-breaking pattern
\begin{equation}
{\rm O}(N) \to {\rm O}(N-1).
\label{vnpat}
\end{equation}
On the other hand, for $v_0>0$ and even $N$ the minimum is realized by
\begin{equation}
\Phi_{\rm min} \sim
\left( 
\begin{array}{l@{\ \ }l@{\ \ }}
I_{N/2}& \;\; 0 \\
\;\; 0 & -I_{N/2} \\
\end{array}
\right)
\label{psivn}
\end{equation}
implying the symmetry-breaking pattern
\begin{equation}
{\rm O}(N) \to {\rm O}(N/2)\otimes {\rm O}(N/2).
\label{vppat}
\end{equation}
For $v_0>0$ and odd values of $N$, we have instead
\begin{equation}
\Phi_{\rm min} \sim
\left( 
\begin{array}{l@{\ \ }l@{\ \ }}
I_{(N+1)/2}& \;\; 0 \\
\;\; 0 & - k I_{(N-1)/2} \\
\end{array}
\right),
\label{psivn2}
\end{equation}
$k = (N+1)/(N-1)$, so that 
\begin{equation}
{\rm O}(N) \to {\rm O}(N/2+1/2)\otimes {\rm O}(N/2-1/2).
\label{vppat2}
\end{equation}

\section{RG flow of the ARP$^{N-1}$ LGW theory for  $N\ge 4$}
\label{hopa}

Within the LGW framework, the nature of the transition of ARP$^{N-1}$
models for $N\ge 4$ can be investigated by studying the RG flow of the
$\Phi^4$ theory (\ref{ahlg}) in the two quartic-coupling space.  For
this purpose, we compute the $\beta$ functions of the model in
different schemes and investigate whether they admit common zeroes
that correspond to stable FPs of the RG flow. If a stable FP exists, a
second-order transition is possible. Otherwise, any transition must be
of first order.

\subsection{The ${\overline {\rm MS}}$ perturbative scheme}
\label{msbar}

We compute the $\beta$ functions in the ${\overline {\rm MS}}$
renormalization scheme~\cite{TV-72}, which uses dimensional
regularization around four dimensions, and the modified
minimal-subtraction prescription ~\cite{ZJ-book}.  The ${\overline
  {\rm MS}}$ $\beta$ functions are defined by
\begin{equation}
\beta_u(u,v) = \left. \mu{\partial u\over \partial \mu}\right|_{u_0,v_0},\quad
\beta_v(u,v) = \left. \mu{\partial v\over \partial \mu}\right|_{u_0,v_0},
\label{betafms}
\end{equation}
where $\mu$ is the renormalization energy scale of the $\overline{\rm
  MS}$ scheme. Here, $u$ and $v$ are the renormalized couplings
corresponding to $u_0$, $v_0$, defined so that $u\propto
u_0/\mu^\epsilon$ and $v\propto v_0/\mu^\epsilon$ at the lowest
order. We compute the $\beta$ functions up to five loops. The complete
series for $N=4$ are reported in App.~\ref{hopert}.

The matrix model is equivalent to the O(2) and O(5) $\Phi^4$ theories
for $N=2$ and $N=3$, respectively.  Therefore, the $\beta$ functions
of the matrix model should be related to the $\beta$ function
$\beta_{{\rm O}(n)}(g)$ of the $O(n)$ model. Using Eq.~(\ref{eq2e3})
we obtain
\begin{equation}
\beta_u + {1\over 2}\beta_v  =  \beta_{{\rm O}(n)}(g) 
\label{p2bf}
\end{equation}
where $g=u+v/2$ and $n=2,5$ for $N=2,3$, respectively.
This exact relation provides a stringent check of 
the five-loop series for model (\ref{ahlg}).

\subsubsection{One-loop analysis close to four dimensions}
\label{oneloop}

Let us first analyze the one-loop $\beta$ functions.  They read
\begin{eqnarray}
\beta_u(u,v) &=& -\epsilon u  + {N^2+N + 14\over 12} u^2  \label{betau}\\
&+&{2N^2 + 3N - 6\over 6N} uv
+ {N^2 + 6\over 4 N^2 } v^2,\nonumber \\
\beta_v(u,v) &=& -\epsilon v  + 2uv + {2 N^2+ 9 N -36\over 12 N} v^2 .
\label{betav}
\end{eqnarray}
The normalization of the renormalized variables can be easily read
from these series.

The one-loop $\beta$ functions (\ref{betau}) and (\ref{betav}) have
four different FPs.  Two of them have $v = 0$ and are always unstable.
The first one is the trivial Gaussian FP at $(u=0,v=0)$, which is
always unstable with respect to both quartic perturbations.  There is
also an O($M$) symmetric FP with $M=(N^2+N-2)/2$ at
\begin{equation}
 u=\epsilon \,{12\over N^2+N+14},\quad v=0, 
 \label{onfp}
 \end{equation}
which can be shown, nonperturbatively, to be unstable with respect to
the operator ${\rm Tr}\,\Phi^4$ for any $N\ge 4$.  Indeed, such
operator contains a spin-4 perturbation with respect to the O($M$)
group~\cite{CPV-03}, which is relevant at the O($M$)-symmetric FP for
any $M>4$ to $O(\epsilon)$, and for any $M\ge 3$ in three
dimensions~\cite{CPV-00,HV-11}. The other two FPs, that have both $v <
0$, only exist for $N< N_{c,0} = 3.6242852...$, One of them is stable,
the other is unstable. For $N = N_{c,0}$ these two FPs merge; for $N>N_{c,0}$,
they become complex.

\subsubsection{Five-loop $\epsilon$ expansion analysis}

To understand the behavior of the system for $\epsilon = 1$, we first
determine the fate of the stable FP that exists for $N<N_{c,0}$ close
to four dimensions. For finite $\epsilon$, we expect a stable and an
unstable FP with $v < 0$ up to $N = N_c(\epsilon)$. The two FPs merge
for $N=N_c(\epsilon)$ and become complex for $N > N_c(\epsilon)$.  We
expand $N_c(\epsilon)$ as
\begin{equation}
N_c(\epsilon) = N_{c,0} + \sum_{n=1} N_{c,n} \epsilon^n,
\end{equation}
and require 
\begin{eqnarray}
&& \beta_u(u,v,N_c) = \beta_v(u,v,N_c) = 0,
\nonumber \\
&& {\rm det}\, \Omega(u,v,N_c) = 0,
\end{eqnarray}
where 
$\Omega_{ij} =
\partial \beta_{g_i}/ \partial g_j$ (where $g_{1,2}$ correspond to
$u,v$)  is the stability matrix.
The last equation is a consequence of the coalescence of the 
two FPs at $N = N_c$. A straightforward calculation gives 
\begin{eqnarray}
N_c(\epsilon) &=& 3.62429 - 0.08865 \epsilon + 0.24968 \epsilon^2 -
                0.69870 \epsilon^3 \nonumber \\ 
     && + 2.88754 \epsilon^4 + O(\epsilon^5).
\end{eqnarray}
The expansion alternates in sign, as expected for a Borel-summable
series.  Resummations using Pad\'e-Borel approximants are stable.  We
obtain $N_c(\epsilon = 1) = 3.60(1)$ using the series to order
$\epsilon^3$ and $N_c(\epsilon = 1) = 3.64(1)$ at order $\epsilon^4$
(the number in parentheses indicates how the estimate changes by
varying the resummation parameters). Apparently, $N_c$ varies only
slightly as $\epsilon$ changes from 0 to 1. In particular, this
analysis predicts the absence of stable FPs for any integer $N\ge 4$
in three dimensions.

\subsubsection{High-order analysis in three dimensions}
\label{homsbar}

The analysis based on the $\epsilon$ expansion allows us to find only
the 3D FPs which are the analytic continuation of those that exist
close to four dimensions.  However, there are models in which a 3D FP
does not have a four-dimensional counterpart. This is the case of the
3D Abelian Higgs model, which undergoes a continuous transition
\cite{MHS-02,NK-03}, in agreement with experiments on superconductors
~\cite{GN-94}. This implies the existence of a 3D stable FP, in spite
of the absence of FPs close to four dimensions \cite{HLM-74}.  Other
LGW $\Phi^4$ theories that have a 3D stable FP with no
four-dimensional counterpart are those describing frustrated spin
models with noncollinear order~\cite{CPPV-04,NO-15}, the $^3$He
superfluid transition from the normal to the planar
phase~\cite{DPV-04}, and the chiral transitions of the strong
interactions in the case the U(1)$_A$ anomaly effects are
suppressed~\cite{PV-13,PW-84}. It is therefore essential to perform a
direct study of the 3D flow. This is achieved by an alternative
analysis of the $\overline{\rm MS}$ series: the 3D $\overline{\rm MS}$
scheme without $\epsilon$ expansion~\cite{Dohm-85,SD-89,CPPV-04}. The
RG functions $\beta_{u,v}$ are the $\overline{\rm MS}$ functions.
However, $\epsilon\equiv 4-d$ is no longer considered as a small
quantity, but it is set equal to its physical value ($\epsilon=1$ in
our case) before determining the RG flow.  This provides a well
defined 3D perturbative scheme which allows us to compute universal
quantities, without the need of expanding around
$d=4$~\cite{Dohm-85,SD-89}.

To determine the stable FPs of the RG flow, we compute numerically the
RG trajectories. They are determined by solving the differential
equations
\begin{eqnarray}
&&-\lambda {d u\over d\lambda} = \beta_u[u(\lambda),v(\lambda)],\nonumber\\
&&-\lambda {d v\over d\lambda} = \beta_v[u(\lambda),v(\lambda)],
\label{rgfloweq}
\end{eqnarray}
where $\lambda\in [0,\infty)$, with the initial conditions
\begin{eqnarray} 
&&u(0) = v(0) = 0 ,\nonumber \\
&& \left. {d u\over d\lambda} \right|_{\lambda=0} = s\equiv {u_0\over |v_0|},\qquad
\left. {d v\over d\lambda} \right|_{\lambda=0} = \pm 1, \label{ini-rgflow}
\end{eqnarray}
where $s$ parametrizes the different RG trajectories in terms of the
bare quartic parameters, and the $\pm$ sign corresponds to the RG
flows for positive and negative values of $v_0$.  In our study of the
RG flow we only consider values of the bare couplings which satisfy
Eqs.~(\ref{stabcond1}) and (\ref{stabcond2}).

The perturbative expansions are divergent but Borel summable in a
large region of the renormalized parameters.  They are resummed
exploiting methods that take into account their large-order behavior
(see App.~\ref{sesum}), which is computed by semiclassical (hence,
intrinsically nonperturbative) instanton calculations
\cite{LZ-77,ZJ-book,CPV-00}.

\begin{figure}[tbp]
\includegraphics*[scale=\graphicscale,angle=-90]{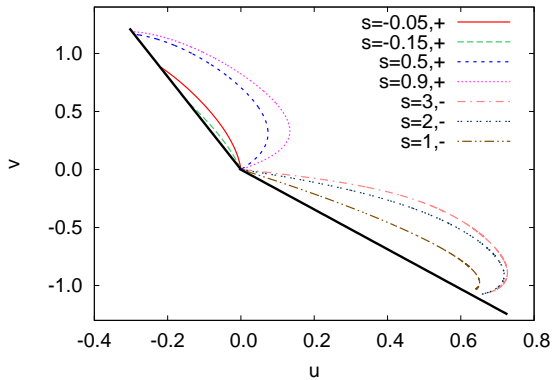}
\caption{(Color online) RG flow of the LGW theory (\ref{ahlg}) for
  $N=4$, in the ${\overline {\rm MS}}$ scheme without $\epsilon$
  expansion, for several values of the ratio $s\equiv u_0/|v_0|$ of
  the bare quartic parameters.  The curves are obtained by solving
  Eqs.~(\ref{rgfloweq}) with the initial conditions
  (\ref{ini-rgflow}): in the legend we report the value of $s$ and the
  sign of $v_0$ ("$+$" and "$-$" correspond to $v_0 > 0$ and $v_0 <
  0$, respectively).  The two solid lines represent the boundary of
  the Borel-summability region, defined by $u + v/4 >0$ and $u + b_4 v
  = u + 7 v/12 > 0$ (see App.~\ref{sesum}).  }
\label{n4trams}
\end{figure}

We present an analysis of the RG flow for $N=4$.  Some RG trajectories
are shown in Fig.~\ref{n4trams}, for several values of the ratio
$s\equiv u_0/|v_0|$.  The RG trajectories flow towards the region in
which the series are no longer Borel summable. In all cases, we do not
have evidence of a stable FP.  These results imply that there is no
universality class characterized by the symmetry breakings
(\ref{vnpat}) and (\ref{vppat}). This would imply a first-order
transition for the ARP$^3$ model.

\subsection{The 3D MZM perturbative scheme}
\label{mzm}

In the massive zero-momentum (MZM)
scheme~\cite{Parisi-80,ZJ-book,PV-02} one performs the perturbative
expansion in powers of the zero-momentum renormalized quartic
couplings directly in three dimensions.  The theory is renormalized by
introducing a set of zero-momentum conditions for the one-particle
irreducible two-point and four-point correlation functions of the
matrix field $\Phi$:
\begin{eqnarray}
&&\Gamma^{(2)}_{a_1a_2,b_1b_2}(p) = 
\left(\delta_{a_1b_2}\delta_{a_2b_1} -{1\over N}
\delta_{a_1a_2}\delta_{b_1b_2} \right)  \times
\label{ren1}  \\
&&\qquad \qquad\times
Z_\phi^{-1} \left[ m^2+p^2+O(p^4)\right],\nonumber \\
&&\Gamma^{(4)}_{a_1a_2,b_1b_2,c_1c_2,d_1d_2}(0) 
= Z_\phi^{-2} m^{4-d} \times
\label{ren2}  \\
&&
\quad \times \left(u U_{a_1a_2,b_1b_2,c_1c_2,d_1d_2} +  
v V_{a_1a_2,b_1b_2,c_1c_2,d_1d_2}\right),
\nonumber 
\end{eqnarray}
where $U,\,V$ are appropriate form factors defined so that $u\propto
u_0/m$ and $v\propto v_0/m$ at the leading tree order.  The FPs of the
theory are given by the common zeroes of the Callan-Symanzik
$\beta$-functions
\begin{equation}
\beta_u(u,v) = \left. m{\partial u\over \partial m}\right|_{u_0,v_0},\quad
\beta_v(u,v) = \left. m{\partial v\over \partial m}\right|_{u_0,v_0}.
\label{betaf}
\end{equation}
The normalization of the zero-momentum quartic variables $u, v$ is
such that their one-loop $\beta$ functions read
\begin{eqnarray}
\beta_u &=& - u  + {N^2 + N + 14\over 18} u^2  
\label{betau0}\\
&+& { 2 N^2 + 3N - 6  \over 9N} uv
+ {N^2 + 6\over 6 N^2} v^2,\nonumber \\
\beta_v &=& - v  + {4\over 3}\,uv 
+ {2N^2 + 9 N - 36 \over 18 N} v^2 .
\label{betav0}
\end{eqnarray}

We compute the MZM perturbative expansions of the $\beta$ functions
and of the critical exponents up to six loops, requiring the
computation of 1428 Feynman diagrams.  The complete expansion for
$N=4$ can be found in App.~\ref{hopert}.  The large-order behaviors of
the series are reported in App.~\ref{sesum}.  The RG trajectories are
obtained by solving differential equations analogous to
Eqs.~(\ref{rgfloweq}) and (\ref{ini-rgflow}).  The $\beta$ functions
are resummed as discussed in Ref.~\cite{ZJ-book,CPV-00} , using the
results of App.~\ref{sesum} for the large-order behavior.  Their
analytic properties close to the FPs are discussed in
Refs.~\cite{nonanbeta,CPV-00}.

\begin{figure}[tbp]
\includegraphics*[scale=\graphicscale,angle=-90]{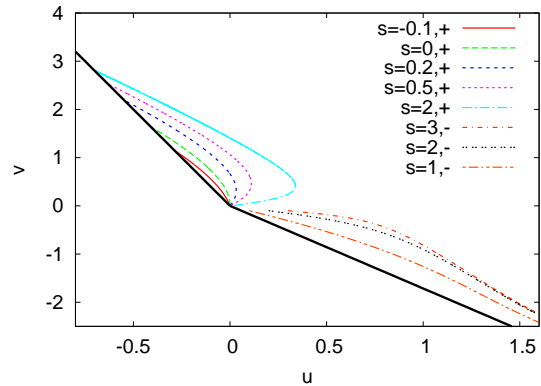}
\caption{(Color online) RG flow of the LGW theory (\ref{ahlg}) for
  $N=4$, in the MZM scheme, for several values of the ratio $s\equiv
  u_0/|v_0|$ of the bare quartic parameters.  The curves are obtained
  by solving Eqs.~(\ref{rgfloweq}) with the initial conditions
  (\ref{ini-rgflow}): in the legend we report the value of $s$ and the
  sign of $v_0$ ("$+$" and "$-$" correspond to $v_0 > 0$ and $v_0 <
  0$, respectively).  The two solid lines represent the boundary of
  the Borel-summability region, defined by $u + v/4 >0$ and $u + b_4 v
  = u + 7 v/12 > 0$ (see App.~\ref{sesum}).  }
\label{n4tramzm}
\end{figure}

Results for $N=4$ are reported in Fig.~\ref{n4tramzm} for several
values of the ratio $s\equiv u_0/|v_0|$.  Most of the RG trajectories
flow towards the region in which the series are no longer Borel
summable. Morever, for $v_0 < 0$ some trajectories flow towards
infinity. In all cases, we do not have evidence of a stable FP,
confirming the analysis in the $\overline{\rm MS}$ scheme.

\section{Numerical results for the ARP$^3$ lattice model}
\label{arp3res}

In this section we present a numerical investigation of the phase
transition of the ARP$^3$ lattice model (\ref{hrpn}).  We set $J=1$.
We present a FSS analysis of MC simulations for cubic $L^3$ systems of
linear size $L$ with periodic boundary conditions.  Because of the
antiferromagnetic nature of the model we take $L$ even. We use a
standard Metropolis algorithm~\cite{footnotemetroarpn}.  We present
results on lattices of size $L\le 100$.  In total, the MC simulations
took approximately 50 years of CPU-time on a single core of a
commercial processor.  Simulations on larger lattices would require a
significantly greater numerical effort or the use of a more effective
updating algorithm, which is not available.

\begin{figure}[tbp]
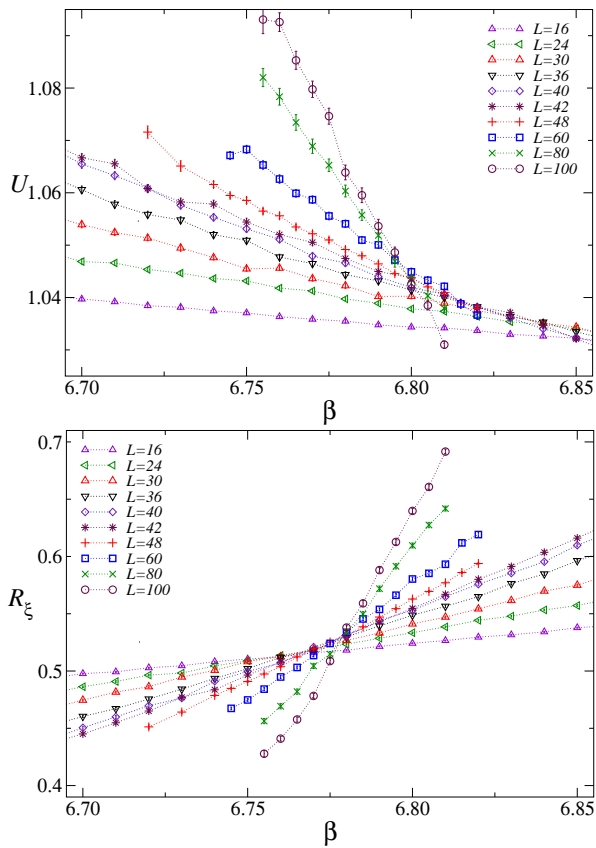

\includegraphics*[scale=\graphicscale]{fig3a.eps}
\includegraphics*[scale=\graphicscale]{fig3b.eps}
\caption{MC estimates of $R_\xi$ (bottom) and $U$ (top) for the ARP$^3$
  lattice model and several lattice sizes $L$ up to $L=100$.  For both
  $R_\xi$ and $U$, the data sets for different $L$ show a crossing
  point for $\beta\approx 6.8$.  The dotted lines are drawn to guide
  the eye.}
\label{rxin4r}
\end{figure}

We compute correlations of the staggered gauge-invariant site variable
$A_{\bm x}^{ab}$, cf. Eq.~(\ref{aiab}). We consider its two-point
correlation function
\begin{equation}
G_A({\bm x}-{\bm y}) = \langle {\rm Tr}\, A_{\bm x}^\dagger  
A_{\bm y} \rangle, 
\label{gxya}
\end{equation}
and, in particular, the corresponding susceptibility and second-moment
correlation length
\begin{eqnarray}
&&\chi =  \sum_{{\bm x}} G_A({\bm x}) = 
\widetilde{G}_A({\bm 0}), 
\label{chisuscara}\\
&&\xi^2 \equiv  {1\over 4 \sin^2 (p_{\rm min}/2)} 
{\widetilde{G}_A({\bm 0}) - \widetilde{G}_A({\bm p})\over 
\widetilde{G}_A({\bm p}},
\label{xidefpbara}
\end{eqnarray}
where ${\bm x}$ runs over lattice points, and ${\bm p} = (p_{\rm
  min},0,0)$ and $p_{\rm min} \equiv 2 \pi/L$.  Moreover, we consider
the quantity
\begin{equation}
U = { \langle [\sum_{{\bm x}} {\rm Tr}\, A_{{\bm 0}}  
A_{\bm x} ]^2 \rangle \over  
\langle \sum_{{\bm x}} {\rm Tr}\, A_{{\bm 0}}  
A_{\bm x} \rangle^2 } ,
\label{binderdefra}
\end{equation}
which is analogous to the so-called Binder parameter.

\begin{figure}[tbp]
\includegraphics*[scale=\graphicscale]{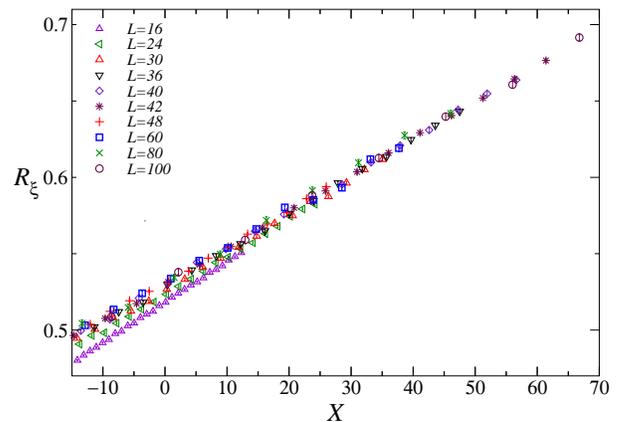}
\caption{ $R_\xi$ versus $X\equiv L^{1/\nu}(\beta-\beta_c)$ with
  $\beta_c=6.779$ and $\nu=0.59$. The data approach a scaling curve
  with increasing $L$, supporting the scaling behavior (\ref{rsca}).
}
\label{rxin4scar}
\end{figure}

To determine the critical behavior we study the finite-size behavior.
The finite-size scaling (FSS) limit is obtained by taking $\beta\to
\beta_c$ and $L\to\infty$ keeping $X \equiv (\beta-\beta_c)L^{1/\nu}$
fixed, where $\beta_c$ is the inverse critical temperature and $\nu$
is the correlation-length exponent. Any RG invariant quantity $R$,
such as $R_\xi\equiv \xi/L$ and $U$, is expected to asymptotically
behave as
\begin{eqnarray}
R(\beta,L) \approx  f_R( X ), \qquad  X\equiv L^{1/\nu}\,(\beta - \beta_c),
\label{rsca}
\end{eqnarray}
where $f_R(X)$ is a universal function apart from a trivial
normalization of the argument. In particular, the quantity $R^* \equiv
f_R(0)$ is universal within the given universality class.  The
corrections to the asymptotic behavior (\ref{rsca}) are expected to
vanish as $L^{-\omega}$ where $\omega>0$ is the universal exponent
associated with the leading irrelevant RG operator.

Fig.~\ref{rxin4r} shows MC data of $R_\xi\equiv \xi/L$ and $U$,
cf. Eqs.~(\ref{xidefpbara}) and (\ref{binderdefra}) respectively, for
several values of $L$. They clearly show a crossing point, providing
evidence of a critical point at $\beta =\beta_c \approx 6.8$.

In order to determine the location and the universal quantities of the
transition, we perform nonlinear fits of $R_\xi$ around
the crossing point. We use the simple Ansatz
\begin{equation}
R_\xi = R_\xi^* + c\,X,
\label{rxisa}
\end{equation}
which should be valid when sufficiently restricting the allowed region
of $\beta$-values around $\beta_c$.
The quality of the fits of $R_\xi$ is reasonably good.  The linear
parametrization (\ref{rxisa}) describes well the data in a relatively
large interval around the transition point, essentially when $\Delta
\equiv |R_\xi-R_\xi^*|/R_\xi^* \lesssim 0.1 $.  We have also performed
fits considering a second-order and a third-order polynomial in $X$,
i.e., fitting $R$ to
\begin{equation}
R = R^* + \sum_{k=1}^n c_k X^k,
\label{fitr}
\end{equation}
with $n=2$ and $n=3$, obtaining consistent results.  The data are not
sufficiently precise to allow us to include scaling corrections in the
fit. Therefore, to estimate their relevance, we have repeated all fits
several times, each time only including data satisfying $L \ge L_{\rm
  min}$, varying $L_{\rm min}$.

We obtain the estimates 
\begin{equation}
\beta_c=6.779(2),\qquad  \nu = 0.59(5), 
\label{fitresr}
\end{equation}
and $R_\xi^* = 0.530(5)$.  The errors the quote are obtained by taking
into account how the results vary when the interval of $\beta$-values
around $\beta_c$ and the minimum size $L_{\rm min}$ are changed.
Statistical errors are significantly smaller.  A scaling plot of
$R_\xi$ is shown in Fig.~\ref{rxin4scar}.  Scaling corrections are
larger for $\beta<\beta_c$ and indeed, the fits are more stable when
only data such that $\beta\gtrsim \beta_c$ are included.

The Binder parameter is much less reliable. As it can be seen from
Fig.~\ref{rxin4r}, the crossing point shows a significant $L$
dependence, indicating the presence of sizeable scaling
corrections. We have performed fits analogous to those performed for
$R_\xi$. We find a significant $L_{\rm min}$ dependence of the
estimates of $\beta_c$, which however appear to converge to the
estimate (\ref{fitresr}) as the size cutoff increases. The estimates
of $\nu$ are consistent with that reported in Eq.~(\ref{fitresr}). As
for the value of of the parameter at the crossing point we find
$U^*\approx 1.04$. We also tried to include scaling corrections. These
fits are however unstable, providing estimates of the
scaling-correction exponent $\omega$ that wildly change with the size
cutoff $L_{\rm min}$.

\begin{figure}[tbp]
\includegraphics*[scale=\graphicscale]{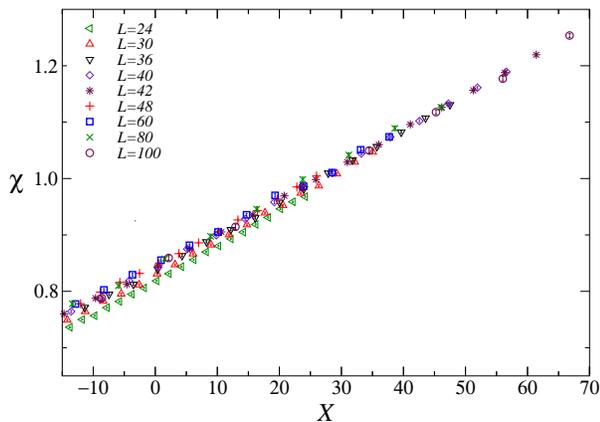}
\caption{ Plot of $\chi/L^{2-\eta}$ versus $X\equiv
  L^{1/\nu}(\beta-\beta_c)$, using $\beta_c=6.779$,
  $\nu=0.59$, and $\eta=0.08$.  
}
\label{chin4scar}
\end{figure}

\begin{figure}[tbp]
\includegraphics[width=8cm]{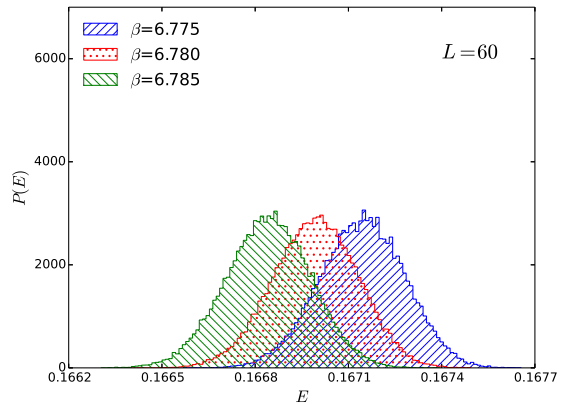}
\includegraphics[width=8cm]{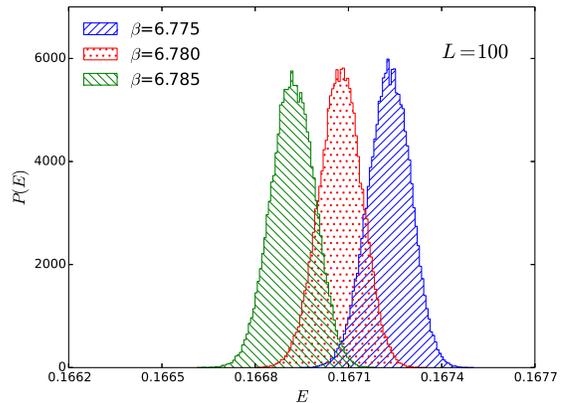}
\caption{Histograms of the energy density, defined as $E=\langle
  H_{\rm RP}\rangle/L^3$, for three values of $\beta$ close to the
  critical point, for $L=60$ (top) and $L=100$ (bottom).  To favor
  comparison, in both figures we use the same range of values of $E$,
  the same energy step, $\Delta E = 10^{-5}$, and a similar number of
  energy measurements, of the order of $10^5$.  There is no evidence
  of double peaks. Moreover, the width of the distributions decreases
  as $L$ increases, as expected at a continuous transition.  }
\label{histos}
\end{figure}

In order to estimate the exponent $\eta$, controlling the spatial
decay of the two-point function $G(x)\sim |x|^{-1-\eta}$ at the
critical point, we analyze the FSS behavior of the susceptibility,
which is expected to be
\begin{equation}
\chi \approx L^{2-\eta} f_\chi(X). 
\label{chisca}
\end{equation}
A fit of the data using the estimates (\ref{fitresr}) gives $\eta =
0.08(4)$, where the error takes also into account the uncertainty on
$\beta_c$ and $\nu$. The corresponding scaling plot is reported in
Fig.~\ref{chin4scar}.

We also mention that analogous results are obtained by considering
observables defined from the two-point function of the gauge invariant
operators $P_{\bm x}^{ab}$, cf. Eq.~(\ref{pdef}), i.e.  $G_P({\bm
  x}-{\bm y}) = \langle {\rm Tr}\, P_{\bm x}^\dagger P_{\bm y}
\rangle$.  The staggered nature of the order parameter is taken into
account be considering correlations only between even points, i.e.,
those such that $p_{\bm x} = (-1)^{\sum_k x_k}=1$.

This numerical study of the ARP$^3$ lattice model provides a robust
evidence that it undergoes a transition at a finite value of
$\beta$.The obtained estimate of $\nu$ also allows us to exclude that
the transition is of first order.  Indeed, at a first-order transition
FSS holds with $\nu=1/d=1/3$~\cite{NN-75,FB-82,PF-83}, while the
estimate (\ref{fitresr}) of $\nu$ is definitely larger than $1/3$.  To
further confirm the continuous nature of the transition, we have also
analyzed the distribution of the energy density, see
Fig.~\ref{histos}. There is no evidence of two peaks and moreover, the
width of the distributions decreases as $L$ increases, as expected at
a continuous transition.  Therefore, we conclude that the ARP$^3$
lattice model undergoes a continuous transition, contradicting the
predictions of the LGW theory.

It may be interesting to compare the estimate (\ref{fitresr}) of the
correlation-length exponent $\nu$ with those of the 3D O($M$) vector
models, which are $\nu=0.629971(4)$ for the Ising ($M=1$) universality
class~\cite{KPSV-16,Hasenbusch-10,CPRV-02,KP-17,GZ-98},
$\nu=0.6717(1)$ for the XY ($M=2$) universality
class~\cite{CHPV-06,KPSV-16,KP-17,GZ-98}, $\nu= 0.7117(5)$ for the
Heisenberg ($M=3$) universality class~\cite{HV-11,CHPRV-02,GZ-98},
$\nu=0.749(2)$ for the O(4) universality
class~\cite{Hasenbusch-01,GZ-98}, $\nu=0.779(3)$ for the O(5)
universality class~\cite{HPV-05,FMSTV-05}, and $\nu \approx 1 - c/M$
with $c=32/(3\pi^2)$ for large $M$~\cite{ZJ-book}. Our results are
consistent with an Ising behavior. However, we do not have any
theoretical argument for this identification, although we note that,
at the transition, there is a breaking of the ${\mathbb Z}_2$ symmetry
associated with the exchange of the even and odd sublattices.

\section{Conclusions}
\label{conclu}

In this work we have studied the critical properties of the 3D
antiferromagnetic RP$^{N-1}$ model, which is characterized by a global
O($N$) symmetry and a discrete ${\mathbb Z}_2$ gauge symmetry.  For
this purpose we present field-theoretical perturbative calculations
and extensive MC simulations.

In the LGW approach one first identifies the order parameter $\Phi$,
then considers the most general $\Phi^4$ theory with the same
symmetries as the original model, and finally determines the stable
fixed points of the RG flow.  If they correspond to a bare theory with
the correct symmetry-breaking pattern, they characterize the possibly
present continuous transitions.  In the presence of gauge symmetries
the method is usually applied by considering a gauge-invariant order
parameter and a LGW field theory that is invariant under the global
symmetries of the original model. In this LGW effective field theory
the gauge degrees of freedom have been integrated out, implicitly
assuming that they are not relevant for the dynamics of the critical
modes.  As already pointed out in Ref.~\cite{PTV-17}, in some cases
this assumption is not correct and the LGW approach may lead to
erroneous conclusions on the nature of the critical behavior.  For
instance, this is the case of the 3D antiferromagnetic CP$^{N-1}$
model characterized by a U(1) gauge symmetry.  In this paper we show
that also in the case of ARP$^{N-1}$ models, which are invariant under
a discrete gauge symmetry, the LGW approach based on a gauge invariant
order parameter may give incorrect predictions on the critical
behavior.

The LGW field theory of ARP$^{N-1}$ models is constructed using the
staggered gauge-invariant composite operator, defined in
Eq.~(\ref{mdef}).  The LGW Hamiltonian does not present cubic terms
due to the antiferromagnetic nearest-neighbor coupling which gives
rise to an additional global ${\mathbb Z}_2$ symmetry.  For $N=3$, the
LGW approach nicely works: its nontrivial prediction of a symmetry
enlargement of the leading critical behavior from O(3) to O(5) has
been accurately verified
numerically~\cite{FMSTV-05,ACFJMRT-05,Carmona-etal-03}.  However, for
$N=4$, the LGW predictions disagree with the numerical results.  The
analyses of the RG flow using high-order perturbative series
(five-loop series in the ${\overline {\rm MS}}$ renormalization
scheme~\cite{TV-72} and six-loop series in the massive zero-momentum
scheme~\cite{Parisi-80,ZJ-book,PV-02}) do not find any evidence of
stable fixed points.  This implies that any transition should be of
first order.  On the other hand, the numerical FSS analysis that we
present for $N=4$ provides evidence of a continuous transition in the
ARP$^3$ model.  This shows that LGW $\Phi^4$ theories constructed
using a gauge-invariant order-parameter field do not generally capture
the relevant features of the critical dynamics when the system has a
discrete gauge symmetry.

These results are analogous to those reported in Ref.~\cite{PTV-17}
for systems with continuous gauge symmetries. In the presence of gauge
symmetries, the main assumption of the LGW approach, i.e., that the
transition is driven by gauge-invariant modes only, may be incorrect,
so that the corresponding field theory may give erroneous predictions
for the nature of the critical behavior.  Therefore, critical gauge
modes should be included to obtain an effective description of the
critical behavior. For example, this happens in the large-$N$ limit of
CP$^{N-1}$ lattice models~\cite{PTV-17}, whose effective
field-theoretical model is the abelian Higgs model for an
$N$-component complex scalar field coupled to a dynamical U(1) gauge
field~\cite{MZ-03}.  We believe that this point deserves further
investigation.

The above considerations should be relevant for several interesting
phase transitions in complex statistical systems.  In particular we
mention the finite-temperature transition of quantum chromodynamics
(QCD). In the limit of $N_f$ massless quarks, the finite-temperature
transition of QCD is related to the restoring of the chiral symmetry.
The nature of the phase transition has been investigated within the
LGW framework~\cite{PW-84,Wilczek-92,RW-93,PV-13,BPV-03}, assuming
that the relevant order-parameter field is a gauge-invariant quark
operators, thus integrating out the gauge degrees of freedom. The present
results show again that this assumption should not be taken for granted.

\appendix

\begin{widetext}

\section{High-order field-theoretical perturbative expansions}
\label{hopert}

In this appendix we report the FT perturbative series of the $\beta$
functions used in our RG analysis of Sec.~\ref{hopa}.  We only report
results for $N=4$; the perturbative series for other values of $N$ are
available on request.

The five-loop $\beta$ functions in the ${\overline{\rm MS}}$ scheme are 
\begin{eqnarray}
\beta_u(u,v) &=& -\epsilon u+
\frac{17 u^2}{6}+\frac{19 u v}{12}+\frac{11 v^2}{32}
-\frac{41 u^3}{12}-\frac{209 u^2 v}{72}-\frac{217 u v^2}{192}-\frac{7v^3}{32}
\\
&+&\frac{67 u^4 \zeta (3)}{18}+\frac{13931 u^4}{1728} 
+\frac{38}{9} u^3 v 
\zeta(3)+\frac{38551 u^3 v}{3456}+\frac{39}{16} u^2 v^2 \zeta (3)+\frac{51059
u^2 v^2}{6912}+\frac{7}{8} u v^3 \zeta (3)+\frac{44671 u v^3}{18432}
\nonumber\\
&+&\frac{145 v^4 \zeta (3)}{1152}+\frac{116401 v^4}{442368}
-\frac{1405 u^5 \zeta (5)}{54}-\frac{14311 u^5 \zeta (3)}{648}+\frac{1139 \pi^4 u^5}{19440}
-\frac{1429027 u^5}{62208}-\frac{4465}{108} u^4 v \zeta(5)
\nonumber\\
&-&\frac{191273 u^4 v \zeta (3)}{5184}+\frac{6479 \pi ^4 u^4 v}{77760}
-\frac{1261847 u^4 v}{31104}-\frac{42625 u^3 v^2 \zeta(5)}{1296}
-\frac{51007 u^3 v^2 \zeta (3)}{1728}+\frac{7907 \pi ^4 u^3 v^2}{155520}
\nonumber \\
&-&\frac{17163385 u^3 v^2}{497664}-\frac{13555}{864} u^2 v^3\zeta(5)
-\frac{571493 u^2 v^3 \zeta (3)}{41472}+\frac{775 \pi ^4 u^2v^3}{41472}
-\frac{16083989 u^2 v^3}{995328}-\frac{81835 u v^4 \zeta(5)}{20736}
\nonumber \\
&-&\frac{580207 u v^4 \zeta (3)}{165888}+\frac{2657 \pi ^4 uv^4}{552960}
-\frac{6769451 u v^4}{1769472}-\frac{61675 v^5 \zeta(5)}{165888}
-\frac{478109 v^5 \zeta (3)}{1327104}+\frac{12853 \pi ^4 v^5}{19906560}
-\frac{5857907 v^5}{15925248}
\nonumber \\
&+&\frac{164689 u^6 \zeta (7)}{864}+\frac{531683 u^6 \zeta (5)}{2592}
-\frac{625u^6 \zeta(3)^2}{864}+\frac{4204813 u^6 \zeta (3)}{41472}
-\frac{23885 \pi^6 u^6}{326592}-\frac{88421 \pi ^4 u^6}{207360}
+\frac{49245733u^6}{663552}
\nonumber \\
&+&\frac{218785}{576} u^5 v \zeta(7)
+\frac{553945 u^5 v \zeta(5)}{1296}-\frac{14041 u^5 v \zeta(3)^2}{5184}
+\frac{53285101 u^5 v\zeta (3)}{248832}-\frac{276925 \pi ^6 u^5 v}{1959552}
-\frac{801097 \pi^4 u^5 v}{933120}
\nonumber \\
&+&\frac{1971278291 u^5 v}{11943936}
+\frac{194285}{512}u^4 v^2 \zeta (7)+\frac{17624591 u^4 v^2 \zeta(5)}{41472}
-\frac{103801u^4 v^2 \zeta (3)^2}{41472}+\frac{436868381 u^4 v^2 \zeta(3)}{1990656}
\nonumber \\
&-&\frac{222605 \pi ^6 u^4 v^2}{1741824}-\frac{23850529 \pi^4u^4 v^2}{29859840}
+\frac{17020433863 u^4 v^2}{95551488}+\frac{1078049 u^3v^3 \zeta(7)}{4608}
+\frac{10457663 u^3 v^3 \zeta(5)}{41472}
\nonumber \\
&-&\frac{120935 u^3 v^3 \zeta(3)^2}{124416}
+\frac{268220975 u^3v^3 \zeta (3)}{1990656}-\frac{3302315 \pi^6 u^3v^3}{47029248}
-\frac{3326713 \pi ^4 u^3 v^3}{7464960}+\frac{3524497739u^3 v^3}{31850496}
\nonumber \\
&+&\frac{3219251 u^2 v^4 \zeta(7)}{36864}+\frac{20008055 u^2 v^4 \zeta(5)}{221184}
-\frac{262619 u^2 v^4 \zeta(3)^2}{995328}+\frac{1583254841 u^2 v^4 \zeta(3)}{31850496}
-\frac{9173555 \pi ^6 u^2 v^4}{376233984}
\nonumber \\
&-&\frac{25105729\pi^4 u^2 v^4}{159252480}
+\frac{60138923803 u^2 v^4}{1528823808}+\frac{218981 u v^5 \zeta(7)}{12288}
+\frac{3957851 u v^5\zeta(5)}{221184}-\frac{58427 u v^5 \zeta (3)^2}{663552}
\nonumber \\
&+&\frac{639020717 u v^5 \zeta (3)}{63700992}-\frac{1208975 \pi ^6 uv^5}{250822656}
-\frac{390793 \pi ^4 u v^5}{11943936}+\frac{22934142763 u v^5}{3057647616}
+\frac{597163 v^6 \zeta (7)}{393216}
\nonumber \\
&+&\frac{47034553 v^6\zeta(5)}{31850496}
-\frac{17285 v^6 \zeta (3)^2}{884736}+\frac{429887731v^6 \zeta (3)}{509607936}
-\frac{394585 \pi^6v^6}{1003290624}-\frac{507587 \pi^4 v^6}{169869312}
+\frac{4964312347v^6}{8153726976} \; ,
\nonumber
\end{eqnarray}
\begin{eqnarray}
\beta_v(u,v) &=& -\epsilon v
+ 2 u v+\frac{2 v^2}{3}
-\frac{127 u^2 v}{36}-\frac{191 u v^2}{72}-\frac{89 v^3}{192}
\\
&+&
\frac{46}{9} u^3 v \zeta (3)+\frac{5543 u^3 v}{864}+\frac{35}{6} 
u^2 v^2\zeta(3)+\frac{24655 u^2 v^2}{3456}+\frac{53}{24} u v^3 \zeta(3)
+\frac{4705 u v^3}{1536}+\frac{343 v^4 \zeta (3)}{1152}+\frac{29345 v^4}{55296}
\nonumber\\
&-&\frac{1855}{54} u^4 v \zeta (5)-\frac{1199}{48} u^4 v \zeta (3)+\frac{1}{10}\pi^4 u^4 v
-\frac{433597 u^4 v}{20736}-\frac{8525}{162} u^3 v^2 \zeta(5)-\frac{24923}{648} u^3 v^2 \zeta(3)
+\frac{1543 \pi ^4 u^3 v^2}{9720}
\nonumber\\
&-&\frac{2047675 u^3 v^2}{62208}
-\frac{41305 u^2 v^3 \zeta(5)}{1296}-\frac{239593 u^2 v^3 \zeta (3)}{10368}
+\frac{163 \pi ^4 u^2 v^3}{1728}-\frac{10811653 u^2 v^3}{497664}
-\frac{97445 u v^4 \zeta(5)}{10368}
\nonumber\\
&-&\frac{67753 u v^4 \zeta (3)}{10368}
+\frac{577 \pi ^4 uv^4}{23040}-\frac{2278937 u v^4}{331776}-\frac{12335 v^5 \zeta(5)}{10368}
-\frac{82787 v^5 \zeta (3)}{110592}+\frac{6251 \pi^4v^5}{2488320}-\frac{12968855 v^5}{15925248}
\nonumber\\
&+&\frac{4753}{18} u^5 v \zeta (7)+\frac{312229 u^5 v \zeta(5)}{1296}
+\frac{313}{162} u^5 v \zeta (3)^2+\frac{54641}{486} u^5 v\zeta (3)
-\frac{3415 \pi ^6 u^5 v}{30618}-\frac{106439 \pi ^4 u^5 v}{186624}
\nonumber\\
&+&\frac{47092103 u^5 v}{746496}
+\frac{36995}{72} u^4 v^2 \zeta(7)
+\frac{90269}{192} u^4 v^2 \zeta (5)+\frac{8611 u^4 v^2 \zeta(3)^2}{2592}
+\frac{3045691 u^4 v^2 \zeta (3)}{13824}-\frac{215305 \pi^6 u^4 v^2}{979776}
\nonumber\\
&-&\frac{4203457 \pi ^4 u^4 v^2}{3732480}
+\frac{748385887u^4 v^2}{5971968}
+\frac{81389}{192} u^3 v^3 \zeta (7)+\frac{1494113 u^3v^3 \zeta (5)}{3888}
+\frac{47945 u^3 v^3 \zeta(3)^2}{15552}
\nonumber\\
&+&\frac{22766455 u^3 v^3 \zeta(3)}{124416}
-\frac{1048795 \pi^6 u^3 v^3}{5878656}
-\frac{6785357 \pi ^4 u^3v^3}{7464960}
+\frac{336567239 u^3 v^3}{2985984}+\frac{3438967 u^2 v^4\zeta(7)}{18432}
\nonumber\\
&+&\frac{10313795 u^2 v^4 \zeta (5)}{62208}+\frac{966247u^2 v^4 \zeta(3)^2}{497664}
+\frac{39980171 u^2 v^4 \zeta(3)}{497664}-\frac{2026955 \pi ^6 u^2 v^4}{26873856}
-\frac{2808157 \pi^4u^2 v^4}{7464960}
\nonumber\\
&+&\frac{2625960661 u^2 v^4}{47775744}
+\frac{1612541 u v^5\zeta(7)}{36864}
+\frac{37924565 u v^5 \zeta(5)}{995328}
+\frac{683903 uv^5 \zeta (3)^2}{995328}+\frac{9249647 u v^5 \zeta(3)}{497664}
\nonumber\\
&-&\frac{6338845 \pi ^6 u v^5}{376233984}-\frac{19159367 \pi^4u v^5}{238878720}
+\frac{873191227 u v^5}{63700992}
+\frac{71197 v^6 \zeta(7)}{16384}
+\frac{233623 v^6 \zeta (5)}{62208}+\frac{260705 v^6 \zeta(3)^2}{2654208}
\nonumber\\
&+&\frac{57288493 v^6 \zeta (3)}{31850496}-\frac{1614635 \pi^6 v^6}{1003290624}
-\frac{6779963 \pi ^4 v^6}{955514880}+\frac{2063219231v^6}{1528823808} \; .
\nonumber
\end{eqnarray}
The six-loop $\beta$ functions in the MZM scheme are 
\begin{eqnarray}
\beta_u(u,v) &=& -u+ 
{17\over 9}u^2+{19\over 18} u v+{11\over 48} v^2
  -1.02241 u^3-0.86877 u^2 v-0.337791 u v^2-0.0648148 v^3
\qquad \label{mzmbeu}\\
&+& 
1.10201 u^4+1.46237 u^3 v+0.955215 u^2 v^2+0.324785 u v^3+0.038265 v^4
-1.53069 u^5-2.60142 u^4 v
\nonumber\\
&-&2.1847 u^3 v^2 - 1.04534 u^2 v^3-0.254045 uv^4
-0.0237542 v^5
+2.50632 u^6+5.35856 u^5 v+5.73074 u^4 v^2
\nonumber\\
&+&3.64914 u^3 v^3 + 1.35588 u^2v^4+0.270658 u v^5+0.0227444 v^6
-4.72398 u^7-11.9943 u^6 v-15.2682 u^5 v^2
\nonumber\\
&-&11.9206 u^4 v^3
-5.84657 u^3v^4-1.75721 u^2 v^5-0.297545 u v^6-0.0218304 v^7 \; ,
\nonumber
\end{eqnarray}
\begin{eqnarray}
\beta_v(u,v) &=& -v+
{4\over 3} u v+ {4\over 9} v^2
-1.05533 u^2 v-0.794696 u v^2-0.140261 v^3
+1.02151 u^3 v
\label{mzmbev}\\  
&+&1.14231 u^2 v^2+0.467669 u v^3+0.0748728 v^4
-1.62911 u^4 v-2.52369 u^3 v^2-1.60381 u^2 v^3
-0.497092 u v^4
\nonumber \\
&-&0.0617621 v^5
+ 2.62863 u^5 v+5.09733 u^4 v^2+4.34004 u^3 v^3+2.00838 u^2 v^4
+0.491214 uv^5
+0.0495342 v^6
\nonumber \\
&-&5.29153 u^6 v-12.4905 u^5 v^2-13.3773 u^4 v^3-8.20642 u^3 v^4-2.97662 u^2
v^5-0.595112 u v^6-0.0508251 v^7 \; .
\nonumber
\end{eqnarray}
\end{widetext}

\section{Summation of the pertubartive series}
\label{sesum}

Since perturbative expansions are divergent, resummation methods must
be used to obtain meaningful results.  Given a generic quantity
$S(u,v)$ with perturbative expansion $S(u,v)= \sum_{ij} c_{ij} u^i
v^j$, we consider
\begin{equation}
S(x u,x v) = \sum_k s_k(u,v) x^k,
\label{seriesx}
\end{equation}
which must be evaluated at $x=1$. The expansion (\ref{seriesx}) in
powers of $x$ is resummed by using the conformal-mapping method
\cite{ZJ-book} that exploits the knowledge of the large-order behavior
of the coefficients, generally given by
\begin{equation}
s_k(u,v) \sim k! \,[-A(u,v)]^{k}\,k^b\,\left[ 1 + O(k^{-1})\right].
\label{lobh}
\end{equation}
The quantity $A(u,v)$ is related to the singularity $t_s$ of the Borel
transform $B(t)$ that is nearest to the origin: $t_s=-1/A(u,v)$.  The
series is Borel summable for $x > 0$ if $B(t)$ does not have
singularities on the positive real axis, and, in particular, if
$A(u,v)>0$.  The large-order behavior can be determined generalizing 
the discussion presented in Refs.~\cite{LZ-77,ZJ-book}.
For even values of $N$, the expansion is Borel summable for
\begin{equation}
u + b_N v > 0,\qquad u + \frac{1}{N}v > 0,
\label{brr}
\end{equation}
where $b_N$ is given in Eq.~(\ref{stabcond1}).  For odd $N$ we obtain 
analogously 
\begin{equation}
u + b_N v > 0,\qquad u + c_N v > 0,
\label{brr1}
\end{equation}
where $c_N$ is given in Eq.~(\ref{stabcond3}).  
Note that the
conditions for Borel summability on the renormalized couplings
correspond to the stability conditions (\ref{stabcond1}) and
(\ref{stabcond2}) of the bare quartic couplings.  In the
Borel-summability region, for even values of $N$, the coefficient 
$A(u,v)$ is given by
\begin{equation}
A(u,v) = \frac{1}{2} \,{\rm Max} \left( u+b_N v,u+v/N\right).
\label{afg}
\end{equation}
For odd $N$, the same formula holds, replacing $u+v/N$ with $u + c_N v$.
Under the additional assumption that the Borel-transform singularities
lie only in the negative axis, the conformal-mapping method turns the
original expansion into a convergent one in the region (\ref{brr}).
Outside, the expansion is not Borel summable.

In the MZM scheme, the large-order  behavior is still given 
by Eq.~(\ref{lobh}). For even $N$, we have 
\begin{eqnarray}
&&A(u,v) =a \; {\rm Max} \left( u+b_N v,u+v/N\right),
\label{afgmzm}\\
 &&a = 0.14777422...,
 \nonumber
\end{eqnarray}
while, for odd values of $N$, $u+v/N$ should be replaced with
$u + c_N v$.

Resummations are performed employing the conformal-mapping method,
following closely
Refs.~\cite{ZJ-book,CPV-00}.  Resummations depend
on two parameters ($\alpha$ and $b$ in the notations of 
Refs.~\cite{ZJ-book,CPV-00}), which are optimized in the procedure.

\end{document}